\documentclass{article}

\def\abstract#1{\vskip 7mm 
        \begin{center}{\large Abstract}\par \smallskip
                \begin{minipage}[c]{12cm}
                        \small #1
                \end{minipage}
        \end{center}
}
\def\title#1{\begin{center}{\Large\bf #1}\end{center}}
\def\author#1{\vskip 5mm \begin{center}{#1}\end{center}}
\def\address#1{\begin{center}{\it #1}\end{center}}
\makeatletter
\@ifundefined{lesssim}{}{}
\@ifundefined{gtrsim}{}{}
\def\vereq#1#2{\lower3pt\vbox{\baselineskip1.5pt \lineskip1.5pt
\ialign{$\m@th#1\hfill##\hfil$\crcr#2\crcr\sim\crcr}}}
\makeatother

\begin{document}

\title{The Cosmic Quartet
  \smallskip \\
  {\large Cosmological Parameters of a Smoothed Inhomogeneous Spacetime}
}
\author{
  Thomas Buchert$^{a,}$\footnote{E-mail:buchert@theorie.physik.uni-muenchen.de} and
  Mauro Carfora$^{b,}$\footnote{E-mail:mauro.carfora@pv.infn.it}
}
\address{$^a$Theoretische Physik, Ludwig--Maximilians--Universit\"at,
Theresienstr. 37, D--80333 M\"unchen, Germany \\
and Department of Physics and Research Center for the Early Universe
(RESCEU)\\ School of Science, The University of Tokyo, Tokyo 113--0033, Japan \\
\smallskip
$^b$Dipartimento di Fisica Nucleare e Teorica,
Universit\`{a} degli Studi di Pavia \\ and 
Istituto Nazionale di Fisica Nucleare, Sezione di Pavia,
via A. Bassi 6, I--27100 Pavia, Italy
}

\abstract{\smallskip
We discuss the relation  between `bare' cosmological parameters as the true 
spatial average characteristics that determine the cosmological model, and 
the parameters interpreted by observers with a ``Friedmannian bias'', i.e., 
within a homogeneous space geometry. We may say that the latter are `dressed' 
by the smoothed--out geometrical inhomogeneities of the surveyed spatial 
region. We identify two effects that quantify the difference between `bare' 
and `dressed' parameters: `curvature backreaction' and `volume effect'. 
An estimate of the latter is given in terms of a simple geometrical example.
}

\section{Introduction}

Contemporary cosmology is heading, as is the generally held believe, 
towards a theoretical home--stretch marked by the determination of 
three cosmological parameters that characterize the standard model 
of cosmology and with it the whole global history of our Universe.
This standard model looks back on a long period of challenges that 
were mastered by its apparent robustness. Essentially, it is based 
on solutions of Einstein's equations for a homogeneous and isotropic 
distribution of matter in the Universe. So, it is indeed a simple 
model that was iconed by Neta Bahcall et al. \cite{bahcall} in terms of a 
`cosmic triangle'. The final phase of combat, in which we are in at 
present and as a large body of cosmologists would put it, consists 
of a phase of high--precision cosmology which, through a plethora of 
orthogonal observations, will soon exclaim highly accurate numbers 
for those three cosmological parameters. But, should we really 
believe this enormously simplified pledge of the standard model 
en--chaining predictions throughout all space and time?

A more level--headed approach immediately reminds us of the 
foundations of the standard model: it ignores structure in the 
Universe. Of course, nobody is as naive as ignoring the existence 
of structure, which in the recent past was celebrated by a number 
of observational successes: the discovery of large--scale structure, 
essentially blowing up the picture of a homogeneous Universe and 
replacing it by a strongly proncounced honeycomb structure of the 
distribution of galaxies. Those structures are extending as deep as 
hundred millions of lightyears and are still continuing at the 
borders of our maps of the Universe such as that currently drawn 
by the Sloan Digital Sky Survey group. 
The ``excuse'' is a conjecture: to maintain the standard model 
we have to assume that {\em on average} the cosmic evolution follows the 
Friedmannian equations whose solutions are fixed by the cosmic 
triangle.

Guided by the idea that a universe model should remain simple, 
while respecting the inhomogeneities that are present in the matter 
distribution as well as in the geometry of spacetime, we have worked out
a cosmology that can still be 
characterized by `cosmological parameters', but there are essentially 
three important colors that have to be added to the standard picture:
firstly, replacing the parameters of the homogeneous universe model 
by corresponding spatial averages entails a scale--dependence, i.e., 
the values of the parameters must depend on the spatial scale over 
which we average the distribution. Secondly, in addition to the 
standard model parameters, there is a fourth player which encodes 
the fluctuations of matter inhomogeneities \cite{buchert:grgdust}. 
If averaged on sufficiently large scales this new parameter
may be a small quantity, but it adds a substantial qualitative 
impact: cosmological parameters are now allowed to communicate, they 
evolve freely governed by the evolution of structure and they are not 
fixed each individually by initial conditions as in the standard model \cite{bks};
we may say that the `cosmic triangle' is transformed into a `cosmic terzet'. 
Thirdly, and this turns out to be a more deeply--rooted insight: 
cosmological parameters that we would measure and interprete with a 
``Friedmannian bias'' (i.e. in a homogeneous spacetime geometry) are 
`dressed' \cite{klingonletter}, i.e., they acquire correction factors
stemming from the geometrical inhomogeneities. 

Understanding this latter fact took a long and rather challenging 
route of employing a number of results that were obtained by 
mathematicians who work on Riemannian geometry \cite{klingon}, reinforcing
the Ricci flow as a natural candidate for the smoothing of geometry
\cite{hamilton},\cite{chow}. 
We shall demonstrate in this note that `the emperor's new clothes' can be 
seen, and the effect of `dressing' can be calculated. Possibly, the 
values interpreted with a Friedmannian bias are appreciably different
from the true averages that govern the dynamics of spacetime.
Further work must be devoted to quantifying this effect: could we 
eventually dismiss the cosmological constant, which now seems to be an 
unavoidable thigh of the cosmic triangle? It will be exciting to follow 
this new route of understanding the influence of structure of the 
fluctuating geometry of spacetime itself on the evolution of our 
Universe.

\section{Bare and Dressed Cosmological Parameters}

To begin with let us concentrate on the density parameter which measures the actual
matter content in a spatial slice of the Universe on some spatial scale of observation. 
We assume that we are given a suitable split into space and time, i.e., a foliation of
spacetime. Since the determination of the {\em total mass on a regional domain} 
in relativistic cosmology is a subject of
controversy (see, e.g., \cite{bray}), we only make use of the well--defined concepts of
a local energy density $\varrho$ 
(here, we only refer to the restmass density for the matter model `dust'),  
and its spatial Riemannian volume average evaluated on a compact domain, e.g., a 
geodesic ball ${\cal B}_0$, $\langle\varrho\rangle_{{\cal B}_0}:=
\int_{{\cal B}_0}\varrho d\mu_g / V_{{\cal B}_0}$ with the
Riemannian volume element $d\mu_g$ associated with the 3--metric of the hypersurface,
and the ball's volume $V_{{\cal B}_0} = \int_{{\cal B}_0} d\mu_g$.
Using this averager for all scalar variables in a $3+1$ ADM setting, we can investigate
regional averages of all relevant spatial variables such as the density and, e.g., the expansion 
rate $\langle \theta\rangle_{{\cal B}_0} =:3 H_{{\cal B}_0}$, defining an effective
``Hubble--constant'' on the domain of averaging. Hence, the cosmological density 
parameter that determines the true (`bare') source of the effective dynamics can be 
unambiguously defined.  

The averaged Hamiltonian constraint may be cast into a relation among a set of 
{\em regional cosmological parameters} as the following scale--dependent 
functionals \cite{buchert:grgdust}:
\begin{eqnarray}
\label{standardparameters}
\Omega^M_{{\cal B}_0} : = \frac{8\pi G M_{{\cal B}_0}}{3 V_{{\cal B}_0}
H_{{\cal B}_0}^2 }\;\;;\;\;
\Omega^{\Lambda}_{{\cal B}_0}:= \frac{\Lambda}{3 H_{{\cal B}_0}^2 }
\;\;;\;\;
\Omega^{R}_{{\cal B}_0}:= - \frac{\langle{\cal R}\rangle_{{\cal B}_0}}
{6 H_{{\cal B}_0}^2}\;\;;\;\;
\Omega^{{\cal Q}^K}_{{\cal B}_0} := - \frac{{\cal Q}^K_{{\cal B}_0}}
{6 H_{{\cal B}_0}^2}\;\;,\nonumber\\
{\rm obeying}\;{\rm by}\;{\rm construction}\;{\rm the}\;{\rm Hamiltonian}\;{\rm constraint}\;\;\;
\Omega^M_{{\cal B}_0} \;+\; \Omega^{\Lambda}_{{\cal B}_0}\;+\; 
\Omega^R_{{\cal B}_0}\;+\; \Omega^{{\cal Q}^K}_{{\cal B}_0} \;=\; 1 
\;\;,
\label{omegaconstraint}
\end{eqnarray}
with the ball's material mass content $M_{{\cal B}_0}$, the scalar curvature $\cal R$, and
the cosmological constant $\Lambda$.
In contrast to the standard FLRW cosmological parameters there are 
four players. In the FLRW case there is by definition no fluctuating source
(condensed into the ``kinematical backreaction'' ${\cal Q}^K_{{\cal B}_0}$
that is composed of positive--definite expansion and shear fluctuation terms
(see \cite{buchert:grgdust}). Hence, the {\em effective cosmology} as the spatially
averaged model can be determined by a scale--dependent and regional `cosmic quartet' 
\cite{buchert:onaverage} rather than by a global `cosmic triangle' \cite{bahcall}.

Eq.~(\ref{omegaconstraint}) forms the basis of a discussion of cosmological parameters
as they determine the theoretical model. They may not be, however, directly accessible
to observations. Unlike in Newtonian cosmology, where the corresponding averages have 
a similar form \cite{buchert:average}, it is not straightforward to compare the above 
relativistic average model parameters to observational data.
The reason is that the volume--averages contain information on the 
actually present {\sl geometrical inhomogeneities} within the averaging domain.
In contrast, the ``observer's Universe'' is 
described in terms of a Euclidean or constant curvature model.   
Consequently, the common interpretation of observations within the set of 
the standard model parameters, if extended by the backreaction parameter or not,
neglects the geometrical inhomogeneities that
(through the Riemannian volume--average) are hidden in the average
characteristics of the theoretical cosmology.

We have suggested an answer to this problem in \cite{klingon} (see also 
\cite{carfora:RG} for a preliminary attempt). Let us highlight some results. 

According to \cite{klingon}, the picture discussed above strictly
depends on the ratio between two density profiles defined on the averaging
domain: one is naturally associated with the actual matter content of the
gravitational sources, whereas the other is the mass density corresponding
to the matter content in the given region, but now thought of as averaged
over a geometrically smoothed--out domain $\overline{\cal B}$ with homogeneous geometry:
\begin{equation}
\langle \varrho \rangle_{{\cal B}_0} = M_{{\cal B}_0} / V_{{\cal B}_0}\;\;\;;\;\;\;
\langle \varrho \rangle_{\overline{\cal B}} = M_{\overline{\cal B}} / V_{\overline{\cal B}}\;\;.
\label{averagedensities}
\end{equation}
We have implemented a regional smoothing procedure for the geometry of the hypersurface
under the assumption of preservation of the ball's material content. We can infer already
from (\ref{averagedensities}) that the average density measured with a 
``Friedmannian bias'' is dressed by a {\sl volume effect} due to the difference between the 
volume of a smoothed region and the actual volume of the bumpy region.

A further result that explicitly involves the geometrical smoothing flows 
(e.g., the Ricci flow for the metric) is furnished by a relation between the 
constant regional curvature in the smoothed model (e.g., a FLRW domain) and the actual
regional average curvature in the theoretical cosmology:
\begin{equation}
\label{regionalcurvature}
\overline{\cal R}_{\overline{\cal B}} = \langle {\cal R} 
\rangle_{{\cal B}_0} 
( V_{\overline{\cal B}} / V_{{\cal B}_0})^{-2/3}
- {\cal Q}^R_{{\cal B}_0} \;\;, 
\end{equation}
where we have introduced a novel measure for the ``backreaction'' of geometrical
inhomogeneities capturing the deviations from the standard FLRW space section, 
the {\em regional curvature backreaction}:
\begin{equation}
{\cal Q}^R_{{\cal B}_0}:= \int_0^{\infty} d\beta \; \frac{V_{{\cal B}_{\beta}}(\beta)}
{V_{\overline{\cal B}}}\;\left[ 
\frac{1}{3}\langle \left({\cal R}(\beta) - \langle {\cal R}(\beta)
\rangle_{{\cal B}_{\beta}}\right)^2 \rangle_{{\cal B}_{\beta}}
- 2 \langle{\tilde{\cal R}}^{ab}(\beta){\tilde{\cal R}}_{ab}(\beta) 
\rangle_{{\cal B}_{\beta}}\right]\;,
\end{equation}  
with $\tilde{\cal R}_{ab}: ={\cal R}_{ab}-\frac{1}{3}g_{ab}{\cal R}$ being the trace--free
part of the Ricci tensor ${\cal R}_{ab}$ in the hypersurface.
${\cal Q}^R_{{\cal B}_0}$, built from scalar
invariants of the intrinsic curvature,
appears to have a similar form as the ``kinematical backreaction'' term (that was built
from invariants of the extrinsic curvature). It features two positive--definite parts, 
the {\em scalar curvature amplitude fluctuations} and {\em fluctuations in metrical anisotropies}. 
Depending on which part dominates we obtain an under-- or overestimate of the actual 
averaged scalar curvature, respectively.  
$\beta$ parametrizes integral curves of the smoothing flow for the metric,
so that the expression above indeed refers to the explicit form of this flow. 
Notwithstanding, this term may be estimated by the actual curvature fluctuations, since the 
Ricci flow acts in a controllable way such that the maxima of the curvature inhomogeneities
are monotonically decreasing during the deformation.

From  Eq.~(\ref{regionalcurvature}) we can understand the 
physical content of geometrical averaging. It makes transparent that, in the smoothed model, 
the averaged scalar curvature is `dressed' both by the {\em volume effect} mentioned above, 
and by the {\em curvature backreaction effect} itself. The volume effect is expected
precisely in the form occurring in (\ref{regionalcurvature}), if we think of comparing two regions
of distinct volumes, but with the same matter content, in a constant
curvature space. Whereas the backreaction term encodes the deviation of
the averaged scalar curvature from a constant curvature model, e.g., a FLRW
space section.

Correspondingly, an observer with a ``Friedmannian bias'' would interprete
his measurements in terms of the `dressed' cosmological parameters:
\begin{equation}
\overline{\Omega}^M_{\overline{\cal B}}:=\frac{8\pi G M_{\overline{\cal B}}}{3 V_{\overline{\cal B}}
\overline{H}_{\overline{\cal B}}^2 }\;;\;\overline{\Omega}^{\Lambda}_{\overline{\cal B}}:= 
\frac{\Lambda}{3 \overline{H}_{\overline{\cal B}}^2 }\;;\;
\overline{\Omega}^{R}_{\overline{\cal B}}:= 
-\frac{\overline{\cal R}_{\overline{\cal B}}}
{6 \overline{H}_{\overline{\cal B}}^2}\;;\;
\overline{\Omega}^{{\cal Q}^K}_{\overline{\cal B}}:= 
-\frac{\overline{\cal Q}^K_{\overline{\cal B}}}
{6 \overline{H}_{\overline{\cal B}}^2}\;,\;
{\rm obeying}\;\;\;
\overline{\Omega}^M_{\overline{\cal B}} + 
\overline{\Omega}^{\Lambda}_{\overline{\cal B}}+ 
\overline{\Omega}^R_{\overline{\cal B}}+
\overline{\Omega}^{{\cal Q}^K}_{\overline{\cal B}} = 1 \;\;.
\label{omegaconstraintdressed}
\end{equation}
The latter equation follows from our assumption that the 
smoothing procedure requires to respect the Hamiltonian constraint of Einstein's equations. 
(For a more detailed discussion see \cite{klingonletter}.)

\section{The Volume Effect: a Geometrical Example}

Comparing and using the relations (\ref{omegaconstraint}) and (\ref{omegaconstraintdressed}),
we conclude that the problem has been condensed into a 
recipee--type rule of applying results that were obtained on theoretically quite involved grounds.
Notwithstanding, the formulation appears simple and therefore immediately suggests
questions of practical importance.
Concrete numbers are needed in order to persuade workers in 
observational cosmology of the actual quantitative importance of the outlined effects.
However, to give 
concrete numbers needs a substantial effort beyond the currently available methods. One of the 
obstacles to provide immediate estimates is the dependence of the proposed effects on 
generic dynamical models,
which are not well--developed in relativistic cosmology. To employ 
(non--generic) spherically symmetric models
is one way to go, and we are currently conducting research in this direction.

Hellaby \cite{hellaby:volumematching} used spherically symmetric models of clusters
and voids to obtain the error made by estimating average characteristics in comparison with
parameters of a corresponding FLRW cosmology. The notion `corresponding' is defined in the 
sense of {\em volume matching} proposed earlier by Ellis and Stoeger \cite{ellisstoeger}. To compare 
Hellaby's approach with our framework is difficult in detail. 
The volume fraction which would have to be estimated in our framework is $1$ by 
definition, the `curvature backreaction' being the only geometrical effect. 
But, since Hellaby compares scalar average characteristics that involve the 
averaged curvatures and the averaged densities on a given scale of interest by keeping the volumes
the same, the calculated effect still can be used as an indication for the deviations of the
averages due to geometrical differences of the two models. It must be emphasized, however, that
a detailed comparison has to follow from an explicit investigation of our effects, and therefore
Hellaby's quoted numbers (amounting to $10$--$30$\% for typical clusters and
voids) can at best be used as an indication. Moreoever, as Hellaby pointed out
himself, the effect depends on details of the dynamical model which, in the cases he investigated,
are still simple solutions that can be done analytically. Also, as he mentioned, the effect
would strongly depend on the topology of the domain which he chooses to be the simplest.

A complementary aspect of the problem is that we need to analyze realistic models of 
hypersurfaces in order to learn about the global effect on very large scales; 
note that, even if we have realistically
modelled clusters and voids, we would still need to arrange those patterns according to 
knowledge of the distribution of large--scale structure. In this sense the effect calculated 
by Hellaby (loc.cit.) is plausible, but it will not provide the answer on cosmological scales. 
Let us illustrate this comment: researches employing rough--surface models to realistically 
model e.g. the moon's surface for the purpose of better estimates on reflection intensity of 
light from the surface, would work with factors of $40$ for the fraction of the moon's modelled
inhomogeneous surface to the surface of a sphere (Karl Vogler, priv. comm.). 
For cosmological volumes the situation is
more subtle, since different patterns may increase, but they may also decrease the 
volume of the spatial domain. Concerning this problem we are currently investigating a generic
relativistic model based on a relativistic generalization of Zel'dovich's approximation, which 
is a well--known approximate tool in Newtonian cosmology to model generic structure formation
\cite{bks}. 

\smallskip

For the sake of illustration of the effect of the volume scaling on the mass
density, let us consider the following explicit example.
Its construction requires the following steps:

\begin{itemize}

\item[(i)] Take a large Euclidean manifold, say a flat 3--torus $T^{3}$ of \emph{
typical} size $R\times R\times R$.

\item[(ii)] Remove from such a manifold, $k$ Euclidean balls $B_{E}(\frac{\pi }{2}r)$
of radius $\frac{\pi }{2}r$ with $\frac{r}{R}$ small, (the $\frac{\pi }{2}$
is for computational convenience).

\item[(iii)] Glue, in place of the removed$\ $Euclidean balls $\{B_{E}(\frac{\pi }{2
}r)\}$, a corresponding number $k$ of 3--spherical balls $\{B_{S^{3}}(\frac{
\pi }{2}r)\}$, (\emph{i.e.}, balls cut out of the surface of a 3--sphere $
S_{r}^{3}$, if the 3--sphere has radius $r$ then the ball $B_{S^{3}}(\frac{
\pi }{2}r)$ is one hemisphere of $S_{r}^{3}$).

\item[(iv)] Note that such a glueing requires a \emph{small} amount of negative
curvature along the boundary 2--spheres of $B_{E}(\frac{\pi }{2}r)$ in order
to have a smooth interface between the flat geometry of the torus $T^{3}$
and the positive curvature geometry of the 3--spherical balls $\{B_{S^{3}}(
\frac{\pi }{2}r)\}$.

\item[(v)] Denote by $T_{\rm inh}^{3}$ the resulting curved torus,  where the  $
B_{S^{3}}(\frac{\pi }{2}r)$'s correspond to the local geometric
inhomogeneity over a region of radius $\frac{\pi }{2}r$ in a space section
of the \emph{observed} universe, (a sort of naive Swiss--cheese model).

\end{itemize}

Under such assumptions (typically, if $r/R$ is suitably small) it is
reasonable to conjecture that the geometry of $T_{\rm inh}^{3}$ will Ricci flow
towards the geometry of a flat torus $\overline{T^{3}}$ (with ${\rm Vol}(\overline{
T^{3}})={\rm Vol}(T_{\rm inh}^{3})$), and we can estimate the relative density
fraction $\frac{\langle\varrho \rangle_{B_{\rm inh}}}{\langle\varrho \rangle_{B_{E}}}$ in such a
setting. Note that the generic region $B_{S^{3}}(\frac{\pi }{2}r)$
supporting the local inhomogeneity in $T_{\rm inh}^{3}$ corresponds to a
Euclidean ball $B_{E}(\frac{\pi }{2}r)$ in the flat smoothed--out $
\overline{T^{3}}$.

Under the local mass preservation constraint, the actual $\langle\varrho
\rangle_{B_{S^{3}}}$ and the smoothed--out $\langle\varrho \rangle_{B_{E}}$ mass densities are
related to each other by (\ref{averagedensities}):
\begin{equation}
\frac{\langle\varrho \rangle_{B_{S^{3}}}}{\langle\varrho \rangle_{B_{E}}}=
\frac{{\rm Vol}\left( B_{E}(\frac{\pi }{2}r)\right) }{{\rm Vol}
\left( B_{S^{3}}(\frac{\pi }{2}r)\right) }.
\end{equation}
Recall that 
\begin{equation}
{\rm Vol}\left( B_{S^{3}}(\frac{\pi }{2}r)\right) ={\rm Vol}\left( S^{2}(r)\right)
\int_{0}^{\frac{\pi }{2}r}\sin ^{2}\left( \frac{t}{r}\right) dt=\pi^{2}r^{3}\;,
\end{equation}
where ${\rm Vol}\left( S^{2}(r)\right) $ is the area of the 2--sphere of radius $r$. Since 
\begin{equation}
{\rm Vol}\left( B_{E}(\frac{\pi }{2}r)\right) =\frac{4}{3}\pi \left( \frac{\pi }{2}
r\right) ^{3}=\frac{\pi ^{4}}{6}r^{3}\;,
\end{equation}
we get 
\begin{equation}
\frac{\langle\varrho \rangle_{B_{S^{3}}}}{\langle\varrho \rangle_{B_{E}}}=\frac{{\rm Vol}
\left( B_{E}(\frac{\pi }{2}r)\right) }{{\rm Vol}\left( B_{S^{3}}(\frac{\pi }{2}r)\right) }=
\frac{\pi ^{2}}{6}\simeq 1.\,\allowbreak 6449\;.
\end{equation}
This shows that indeed one can have a significant mismatch between the
actual average mass density and the smoothed--out mass density owing only to the volume
effect (the curvature backreaction effect was not considered in this rather qualitative
example).

\section*{Acknowledgments}
TB acknowledges hospitality by the University of Tokyo and financial 
support by the Research Center for the Early Universe (RESCEU, Tokyo),
COE Monkasho Grant. This work is
also partially supported by the Sonderforschungsbereich SFB 375 
`Astroparticle physics' by the German science foundation DFG, and  
by the Ministero dell' Universita' e della Ricerca 
Scientifica under the PRIN project `The Geometry of Integrable Systems'.


\begin{thebibliography}{03}

\bibitem{bahcall}
N.~A. Bahcall, J.~P. Ostriker, S. Perlmutter and P.~J. Steinhardt, Science {\bf 284}, 1481
(1999).

\bibitem{bray}
H.~L. Bray, Notices of the AMS {\bf 49}, 1372 (2002).

\bibitem{buchert:grgdust}
T. Buchert, G.R.G.\ {\bf 32}, 105 (2000).

\bibitem{buchert:onaverage}
T. Buchert, in: {\em 9th JGRG Meeting, 
Hiroshima 1999}, Y. Eriguchi et al. (eds.), pp. 306--321 (2000).

\bibitem{buchert:average}
T. Buchert and J. Ehlers, A.\& A. {\bf 320}, 1 (1997).

\bibitem{bks}
T. Buchert, M. Kerscher and C. Sicka, P.R.D.\ {\bf 62}, 043525 (2000).

\bibitem{klingon}
T. Buchert and M. Carfora, C.Q.G.\ {\bf 19}, 6109 (2002).

\bibitem{klingonletter}
T. Buchert and M. Carfora, P.R.L.\ {\bf 90}, 31101 (2003).

\bibitem{carfora:RG}
M. Carfora and K. Piotrkowska, P.R.D.\ {\bf 52}, 4393 (1995).

\bibitem{chow}
B. Chow, Preprint: math.DG/0211266 (2002).

\bibitem{ellisstoeger}
G.~F.~R. Ellis and W. Stoeger, C.Q.G. \ {\bf 4}, 1697 (1987).

\bibitem{hamilton}
R.~S. Hamilton, in {\em Surveys in Differential Geometry Vol 2}, 
International Press, pp.7--136 (1995).

\bibitem{hellaby:volumematching}
C. Hellaby, G.R.G.\ {\bf 20}, 1203 (1988).

\end{thebibliography}
\end{document}